\documentclass[ammsymb,prl,aps,twocolumn,superscriptaddress,showpacs]{revtex4}
\usepackage{graphicx}

\def \beq{\begin{equation}}
\def \eeq{\end{equation}}
\def \beqar{\begin{eqnarray}}
\def \eeqar{\end{eqnarray}}
\input{tcilatex}

\begin{document}

\title{Comment on ''Self-Segregation versus Clustering in the Evolutionary
Minority Game''}
\author{E. Burgos}
\affiliation{Departamento de F{\'{\i}}sica, Comisi{\'o}n Nacional de Energ{\'\i }a At{\'o}%
mica, Avda. del Libertador 8250,1429 Buenos Aires, Argentina}
\author{Horacio Ceva}
\affiliation{Departamento de F{\'{\i}}sica, Comisi{\'o}n Nacional de Energ{\'\i }a At{\'o}%
mica, Avda. del Libertador 8250,1429 Buenos Aires, Argentina}
\author{R.P.J. Perazzo}
\affiliation{Departamento de F{\'{\i}}sica FCEN,and Centro de Estudios Avanzados,
Universidad de Buenos Aires, Ciudad Universitaria - Pabell{\'o}n 1, 1428
Buenos Aires, Argentina}
\date{\today}
\maketitle


In a recent letter \cite{weisz} it is claimed that the occurrence of self
segregation in the Minority Game (MG) depends upon the prize-to-fine ratio $%
R $. The authors find that self segregation occurs if $R = 1$, but for $R <1
$, clustering takes place by which the probability density function $P(p)$
turns into a symmetric, inverted U thus indicating that the players tend to
have a common $p$ that is closer to 1/2. $P(p)$ also displays an M-shape in
a narrow transition between both regimes. In the previous numerical
experiments in which the effects of $R$ have been investigated \cite{bc}-%
\cite{bcp-II} clustering was not found. We consider that the effect reported
in Ref.\cite{weisz} is mainly due to the updating rule for the probability $p
$ (the player's "gene value") that has been used. This consists \cite%
{private} in updating $p$ by choosing at random a new value for it in the
interval [0,1]. This differs from the \textit{gradual} updating normally
used (see, for instance, Ref.\cite{Johnson}) by which $p$ is updated $%
p\rightarrow p^{\prime }$ with $p^{\prime }$ at random in $[p-\delta
p,p+\delta p]$, with $\delta p\ll 1$.

\begin{figure}[tbp]
\includegraphics[width=9cm,clip]{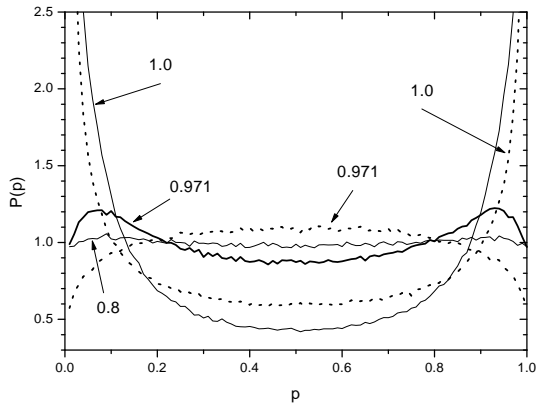}
\caption{Probability functions $P(p)$ obtained for the values of $R$
indicated in the figure. Continuous lines corresponds to the usual
\protect\cite{bcp}, \protect\cite{Johnson} relaxation dynamics with $\protect%
\delta p =0.1$. Dotted lines correspond to the random updating \protect\cite%
{private} of Ref.\protect\cite{weisz} . All curves correspond to $N=10,001$
agents, $10^6$ time steps, and are averaged over 100 histories. In all
cases, the threshold is $d=-4$.}
\label{P(p)}
\end{figure}

The change introduced in Ref.\cite{weisz} amounts to use a different
relaxation dynamics and this is bound to give rise to important changes
\cite {HC}.  The lack of correlations between successive values of $p$,
that is implied in the random updating used in Ref.\cite{weisz}, produces
a result that closely resembles a random walk.  The same result is to be
expected in any circumstance in which $\delta p$ is (or can be) large.
For instance if periodic boundary conditions for $p$ are used
\cite{weisz-III} , it occasionally happens that $\delta p\simeq 1$.  On
the other hand, a gradual updating of $p$ with reflective boundary
conditions for $p$ corresponds\cite{thermal} to the minimization of a cost
function for the ensemble of players.  In Fig.\ref{P(p)} we show how such
new updating rule is the main responsible of the effect reported in
Ref.\cite{weisz}.

In Ref.\cite{weisz} and the related reference \cite{weisz-IV} further
considerations are made concerning the temporal oscillations of $<p>$. The
occurrence of these oscillations are also a consequence of the updating rule
used. In fact such oscillations are not observed for $R<1$ with the usual,
reflective boundary conditions for the updating of $p$. The remaining,
(small) random fluctuations in $<p>$ have been interpreted in terms of a
thermal description \cite{thermal}.

Finally, it is worth to remark that for $R<1 $ the clustering effect
displayed by the departures of $P(p)$ from unity is drastically smaller than
the self segregation effect found for $R>1$, and completely vanishes in the
limit $R\rightarrow 0$

\hspace{2cm}

\end{document}